# Fundamentals of the metal contact to p-type GaN – new multilayer design

Konrad Sakowski[1,2], Paweł Strak[1], Stanislaw Krukowski[1]

[1]*Institute of High Pressure Physics, Polish Academy of Sciences, Sokolowska 29/37, 01-142 Warsaw, Poland*

[2]*Institute of Applied Mathematics and Mechanics, University of Warsaw, 02-097 Warsaw, Poland*

Electrical properties of contact to p-type nitride semiconductor devices, based on gallium nitride were simulated by *ab initio* and by drift-diffusion calculations. The contact electric properties are shown to be dominated by electron transfer form metal to GaN related to Fermi level difference both by *ab initio* and model calculation. The results indicate on high potential barrier for holes leading to nonohmic character of the contact. The electrical nature of the Ni-Au contact formed by annealing in oxygen atmosphere is elucidated. The doping influence on the potential profile in p-type GaN was calculated by in drift-diffusion model. The energy barrier height and width for hole transport is determined. Based on these results,  new type of the contact, is proposed. The contact is created employing multiple layer implantation of the deep acceptors. The implementation of such design promise to attain superior characteristics (resistance) as compared to other contacts used in bipolar nitride semiconductor devices. The development of such contact will remove one of the main obstacles in the development of highly efficient nitride optoelectronic devices both LEDs and LDs: energy loss and the excessive heat production close to the multiple quantum wells system.

I.   **Introduction**

Despite successful development of nitride optoelectronic devices, such as light emitting diodes (LEDs), awarded Nobel Prize in 2014 [1], and laser diodes (LDs) [2], both visible and UV, several formidable obstacles still remain on the path to the considerable improvement of these devices which is still badly needed. The most harmful problems are related to nitrides p-type properties, namely high resistivity of both the plain bulk and also the contact. The first problem was related to absence of shallow acceptor, as the only reliable defect is Ga substitutional Mg atom which creates deep acceptor of the ionization energy of about 170 meV [3]. Therefore at room temperature about 1% of Mg is ionized and p-type conductivity suffers from low density and mobility. Nevertheless, the effective method to overcome the first obstacle was developed, first by the concept of the type doping [4] and subsequently the discovery of the method to obtain mobile hole charge [5]. The methodology promises to obtain temperature-independent hole density, applicable for UV devices [6,7]. As the above solution has been proved to be working, the progress in this respect is a matter of technology.

The second problem has not been solved so far. Its solution is subject of this publication. High contact resistance is related to wide bandgap of the nitrides, which necessarily locates their valence bands (VB) much below the Fermi energy of any metal candidate for the contact. Typically, work-functions of the metal are close to 5 eV, the highest value is reported for gold which is $\phi_{Au} = 5.10 \div 5.47\ eV$ [8]. The data span is related to work-function dependence on the configuration of atoms at the surface of the material. The other metals have lower work-function values, typically in the range $\phi_{Au} = 4.0 \div 5.0\ eV$ [8]. Therefore, the most frequently used metal contacts are based on combination of Au and other metals: Au/Ni [9-11], Pt/Ni/Au [12], Ti/Pt/Au [13], Pd/Au [14].

The work-function difference, or equivalently the Fermi level difference between the metal and the semiconductor is the energy cost at which the holes from the metal can be transferred to the semiconductor at Fermi level. In case of p-type this is close to the to the energy of valence band maximum (VBM), i.e. the ionization potential *I*. Aluminum and gallium nitrides are characterized by strong bonds and low energy of valence band states. Therefore the ionization potentials *I* calculated for metal-terminated clean AlN and GaN are $I_{AlN} = 9.13\ eV$ and $I_{GaN} = 7.54\ eV$ [15]. The ionization potential could be affected by band bending, nevertheless the magnitude of the bending rarely exceeds 1 eV. Therefore, most of this energy difference remains, and it creates the energy barrier $\Delta E$ for holes in excess of $\Delta V_o \cong 2\ eV$. Hence, the direct deposition of the metals at p-type GaN could not be used for construction of any working nitride optoelectronic device. To alleviate this problem, the sophisticated procedures of the contact formation were developed, such as deposition of Ni and Au layers and annealing in oxygen atmosphere. It is used in majority of the present day nitride diodes [9-11,16-20].

In addition to that, the current-voltage (I-V) electric measurement results obtained for LEDs and LDs at high currents, controlled mostly by the contact resistance, exhibit nonlinear behavior, indicating nonohmic type of resistance. Thus, the contact is characterized by the Schottky energy barrier [21]. The detailed investigations revealed the nature of the Ni/Au contact formation on the metal side [9-11, 16-20]. During the annealing process nickel attoms diffuse across gold layer and reacts with oxygen creating NiO layer on the external side. Thus, the chemical potential difference driving the Ni current to the surface is maintained. Consequently, the underneath Ni layer becomes very thin and nonuniform. It is speculated that some of Ga atoms diffuse from GaN to Ni layer. In the result the contact has optimally acceptable resistance of order of $\rho \sim 10^{-4} \Omega\ cm^2$ at the current density $j = 4000\ A\ cm^{-2}$. The claimed minimal voltage at the contact is $\Delta V = 0.4\ V$ [10]. Unfortunately this is not a standard, the results shows considerable scattering in a function of weakly controlled parameters, such as the surface morphology or dislocation density, which most likely induces strain and the instability of the structure. Generally the voltage is much higher, the probable value is $\Delta V \cong 1 \div 3\ V$.

## II. The model

The electric stability of the metal and the nitride heterointerface leads to attainment of the common Fermi energy level [22]. This occurs via increase of the energy of quantum states in the nitride bulk due to creation of the electric dipole layer. The dipole negative part is due to electron transfer to the nitride and the opposite positive part appears in the metal due to electron absence. The screening length in the metal is about $0.01$ Å therefore the effect on the potential energy is negligible [23]. At the surface of contact during the charge transfer the difference of the energy of quantum states of the semiconductor and the metal is preserved, that allows to estimated magnitude of the band bending. Band bending is equivalent to the barrier for the introduction of holes into the semiconductor, which is:

$$E_{bar} \cong E_F(m) - E_F(GaN) \tag{1}$$

where $E_F(m)$ and $E_F(GaN)$ are Fermi energies in the metal and p-type GaN bulk respectively. The large magnitude of this barrier explains the large resistance of as deposited metal contacts, including gold.

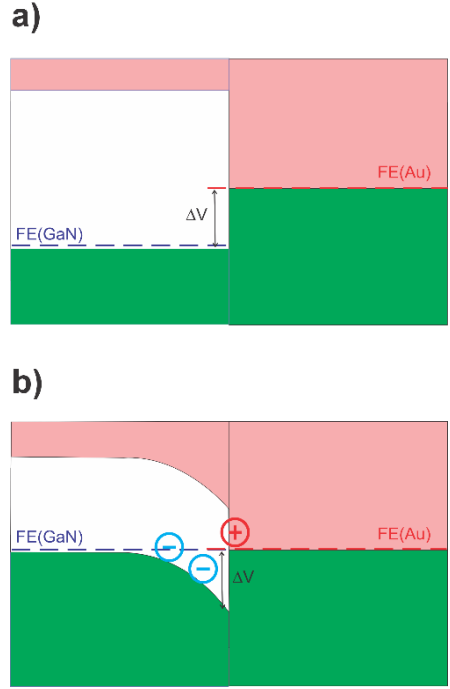

*Fig 1. The band and electric potential profiles of metal – p-type nitride heterointerface: (a) metal-nitride contact without charge transfer, (b) equilibrium profile of the contact resulting in creation of dipole layer. The green and rosy colors represent occupied and empty band states, the white color - absence of the states in the gap.*

These profiles provide explanation of the relative success of oxygen atmosphere annealed Ni-Au contact. In addition to the metal side formation, the outdiffusion of Ga atoms takes place creating layer of high density of the Ga vacancies ($V_{Ga}$). Gallium vacancies are multiple-ionized state deep acceptors, of three ionization energies ($E_A(V_{Ga}^{1-}) \cong 2.35\ eV$, $E_A(V_{Ga}^{2-}) = 1.02 \div 0.62\ eV$, and $E_A(V_{Ga}^{3-}) \cong 0.48\ eV$) [24,25]. In addition, nickel atoms may diffuse over the network of gallium vacancies creating substitutional $Ni_{Ga}$ atom defects, also good candidates for deep acceptors ($E_A(Ni) \cong 1.6 \div 1.5\ eV$) [26,27]. The mixture of these states serve as tunneling channel for holes. It is also likely that the arrangement of Ni layer under Au layer may attain the structure favorable for incorporation of Ga atoms thus facilitating creation of such defect layer.

### III. Calculation procedure

In the determination of the properties of composite Au-GaN system a combination of the various approached were used. First *ab initio* simulations were used to determine basic electronic properties of the separate Au and GaN and the connected Au-GaN system. For these simulation DFT based SIESTA package was used. SIESTA shareware package was developed in Spanish Initiative for Electronic Simulations with Thousands of Atoms (SIESTA)[27]. The fundamental routine solves set of nonlinear Kohn-Sham equations iteratively to obtain set of eigenfunctions and real eigenvalues [27]. The eigenfunctions are a linear combinations of the of radial numeric atomic orbitals of finite extent, multiplied by spherical harmonics, effectively limited to *s, p* and *d* polynomials of angular sine and cosine functions [27-29]. These *s* and *p* orbitals centered on both Ga and N atoms are expressed by triple zeta functions. Gallium *d* shell electrons are incorporated into the valence set. Ga *d* orbitals are single zeta functions. Au basis include 6s, 6p – triple zeta function and 5d, 5f – double zeta [30,31]. The eigenfunctional set is limited to single period of these orbitals using Troullier-Martins atomic pseudopotentials [32,33]. K-space integration is approximated by summation over grid of Monkhorst-Pack points, the number of them depends on the system considered, $(7 \times 7 \times 7)$ and $(5 \times 5 \times 1)$, for Au and Ga bulk and for Au-GaN slab, respectively [34]. In addition GGA-PBE (PBEJsJrLO) functional is parameterized using β, μ, and κ values set by the jellium surface (Js), jellium response (Jr), and Lieb-Oxford bound (LO) criteria [35,36]. The iteration SCF loop was terminated when the difference for any element of the density matrix in two consecutive iterations was below $10^{-4}$.

The *ab initio* calculation for the bulk GaN gave its lattice parameters a: $a_{GaN}^{DFT} = 3.21$ Å and $c_{GaN}^{DFT} = 5.23$ Å close to the experimental values: $a_{GaN}^{exp} = 3.189$ Å and $c_{GaN}^{exp} = 5.186$ Å [37]. The DFT obtained lattice parameter of bulk Au fcc cubic lattice was $a_{Au}^{DFT} = 4.14$ Å, quite close to experimental value $a_{Au}^{exp} = 4.0782$ Å [38]. The quantum states energies, were corrected using Ferreira et al. approach, denoted as GGA-1/2, giving reasonably good band parameters [39,40], e.g. GaN bandgap $E_g^{DFT}(GaN) = 3.47\ eV$, identical to the value from low temperature measurement: $E_g^{exp}(GaN) = 3.47\ eV$ [41,42]. Born-Oppenheimer approximation was applied in all atomic position relaxation. The atom positions relaxation was terminated for the values of forces acting on any atom dropped down below 0.005 eV/Å.

IV. **The calculation results**
   a. **Doping in the GaN bulk**

The creation of such layer is weakly controlled, that explain scattering and unreliability of the contact formation procedure. Naturally, some modifications were introduced, such as In-rich GaInN contact layer. This may be beneficial as due to much lower bandgap, the barrier energy may be smaller. Nevertheless, the basic electric structure remains the same. In all these implementations p-type doping is imposed with high concentration of Mg acceptors, which have the level located at about $E_A = 180\ meV$, above VBM [3]. The Mg doping may be controlled during growth. Typically this affects the position of the Fermi level as only part of the acceptors is ionized. The change of the Fermi level and the concentration of holes are presented in Fig. 2.

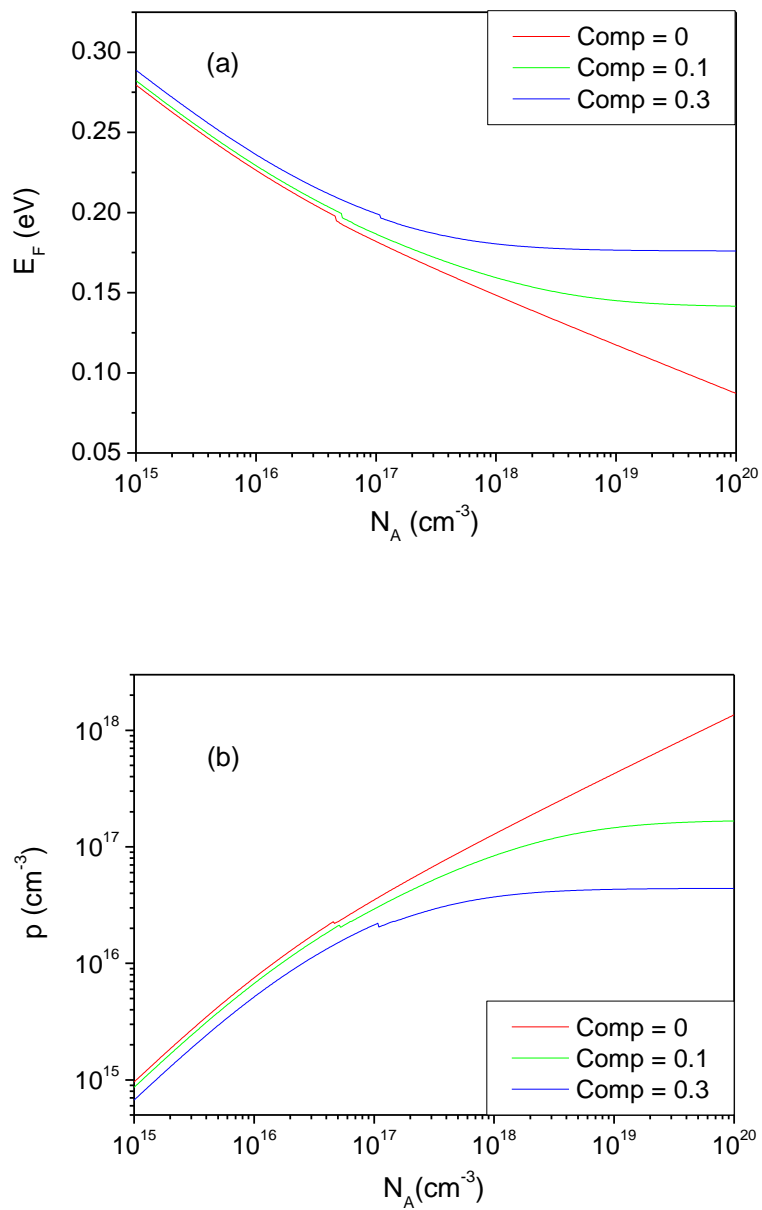

*Fig. 2. Neutrality condition for Mg doping for three values of the compensation: (a) position of Fermi level, (b) hole concentration in function of the concentration of acceptors.*

As it follows, the charge-neutrality Fermi level corresponding to acceptor level $E_A = 180\ meV$ depends on the compensation also. The position of Fermi level at the acceptor energy is attained for $N_A = 1.15 \times 10^{17}\ cm^{-3}$ for absence of compensation. That corresponds to the following density of charged acceptors $N_A^- = 3.81 \times 10^{16}\ cm^{-3}$ (in this case this is the density of holes also). For compensation at 10% and 30% this condition is attained for $N_A = 1.62 \times 10^{17}\ cm^{-3}$ and $N_A = 1.15 \times 10^{18}\ cm^{-3}$, respectively. This this is scaled with respect to the $N_A - N_A$ difference. For lower and higher Mg concentrations, the Fermi level is above and below acceptor energy, respectively. This has drastic influence on the concentration of holes, and accordingly for charged acceptors $N_A^-$ thus affecting the overall functionality of p-type contact as it will be shown below.

### b. Ab initio data

*Ab initio* simulations cannot be used for direct calculation of the semiconductor-metal heterojunctions because the number of the atoms needed for simulations would be prohibitively large reaching tens of thousands of atoms. Hence, the simulations of some features was implemented. Following this route, the electronic properties of bulk GaN and Au systems were obtained from DFT procedure separately. The simulation cells and results are presented in Fig. 3.

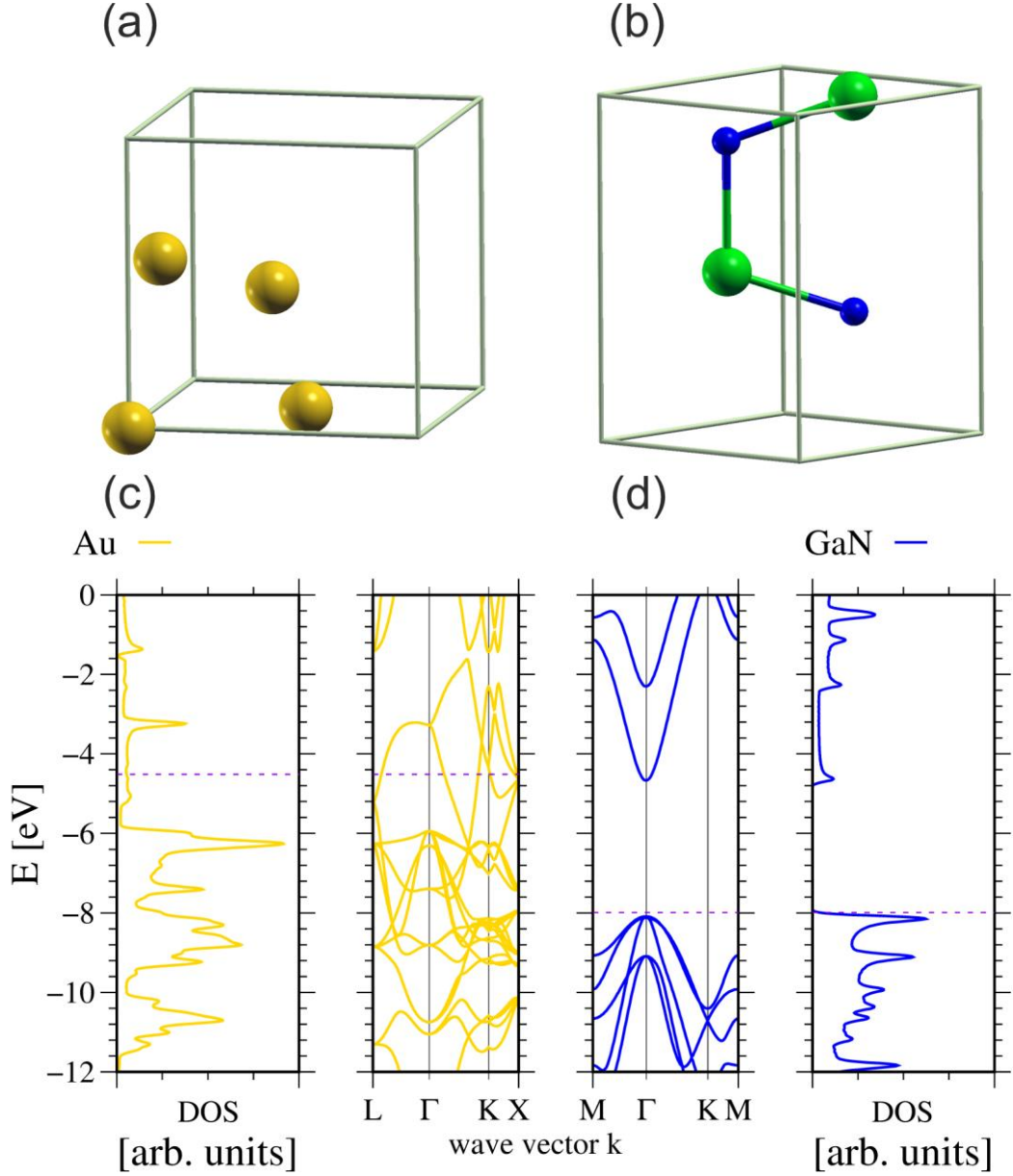

*Fig. 3. Simulation of bulk Au (cubic) and GaN (wurtzite): (a) Au simulation cell (cubic), (b) GaN simulation cell (wurtzite), (c) electronic properties of bulk Au: density of states (DOS) and bond diagram, (d) electronic properties of bulk GaN: density of states (DOS) and bond diagram. The yellow, green and blue balls represent gold, gallium and nitrogen atoms, respectively. The doping level of p-type GaN was set to $N_A = 10^{19}\ cm^{-3}$.*

The calculated electronic properties of both bulk systems are presented in Fig. 3. The panel arrangement was adjusted to facilitate direct comparison of these properties. In fact these properties were calculated and presented in many publications, e.g. GaN [43-45] or Au [46]. The obtained band gap for GaN is direct, equal to $E_g^{DFT}(GaN) = 3.47\ eV$. Additionally, density of states (DOS) of both systems was plotted in the same energy range. For gold DOS has strong maximum at $E_{DOS}^{DFT}(Au) = -6.26\ eV$, associated with the several band maxima at this energy. Similar occurrence is identified for GaN at valence band maximum (VBM) which is located at $E_{DOS}^{DFT}(GaN) = -8.15\ eV$. Important parameter is the Fermi level position, which is affected in case of GaN by incorporation of the acceptor density in SIESTA background simulation procedure to $N_A = 10^{19}\ cm^{-3}$ [27]. From the obtained simulation data it follows that the Fermi level for gold is $E_F(Au) = -4.520\ eV$ and for p-type GaN it is $E_F(p-GaN) = -7.997\ eV$. Therefore the Fermi energy difference is $\Delta E_F = 3.477\ eV$. Creation of heterostructure leads to electron flow from Au to GaN and the downward shift of the energy level in Au and upward shift of the energy level in GaN. Partially this could be observed in the *ab initio* calculations of the GaN-Au slab containing finite number of atomic layers of both crystals. The simulation slab representing Au-GaN heterointerface is presented in Fig. 4. The slab consist of two atomic layers of Au and eight double Ga-N atomic layers. At the bottom it is terminated by hydrogen pseudoatoms with fractional $(Z = 3/4)$ charge.

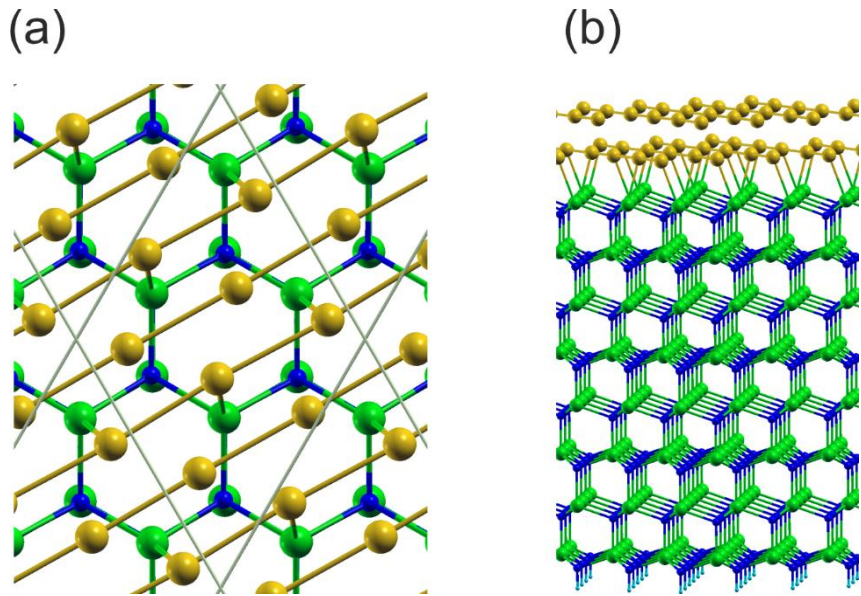

*Fig. 4. Slab used for simulation of Au-GaN heterointerface: (a) – top view, (b) side view. The atoms are denoted as in Fig 3.*

As it is shown, the contact Au atom layer is distorted to adjust to GaN lattice. The topmost Ga-N layer distortion is very small. Nevertheless these diagram indicate that the connection is created and Au is wetting Ga-terminated GaN surface. This confirms good properties of GaN-Au contact observed in experiment.

The electronic properties of Au-GaN heterointerface are presented in Fig 5. As it is shown the properties of the system are determined by the electron transfer from the metal into the semiconductor. The Fermi level of the composite calculated system is determined by the balance of the electrons in Au and p-type GaN. The doping level which was selected relatively high, about $N_A = 3 \times 10^{20}\ cm^{-3}$, generates large number of the empty states in GaN valence band. Therefore the electrons can be relatively easily shifted from the metal creating the potential difference. This is visible from the averaged potential profile shown in Fig. 6.

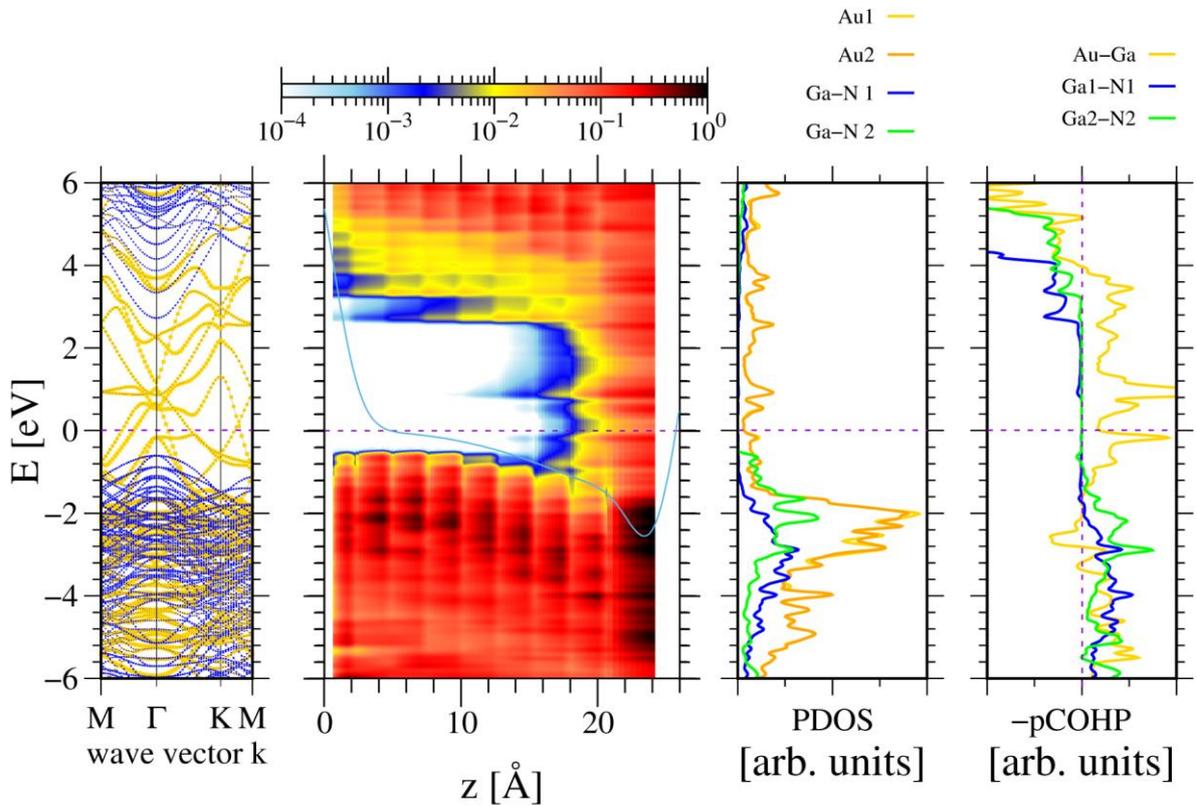

*Fig. 6. Electronic properties of the Au-GaN slab representing Au-GaN heterointerface. The panels present (from the left); band diagram in real and momentum space and projected DOS (PDOS) and Crystal Orbital Hamilton Population [47,48]. The gold atoms Au1 and Au2 are located at the interface and in the second layer, respectively. The pair of Ga-N atoms, Ga-N1 and Ga-N2, are located in the slab second layer next to the Au and H hydrogen pseudoatoms, respectively. The states in momentum space that denote the majority contribution to Au and Ga-N atoms respectively. The blue solid line in real space diagram represent the electron*

*energy in the averaged electrostatic potential, according to the method from Ref. 45 [45]. The line represent the electron energy, i.e. potential profiled multiplied by electron charge q $E_e = qV$, where $q = -e$, and e is elementary charge $e = 1.602 \times 10^{-9}$ C. The doping level of p-type GaN was set to $N_A = 3 \times 10^{20}$ $cm^{-3}$.*

In this case the electrons can be shifted into this region, as it is shown in the diagram. The electron energy profiles shows minimum in the gold region which is related to the electron shift to GaN and net positive charge of Au ions. This leads to the emergence of electric dipole layer in which the GaN states are shifted upwards and the Au states downwards. This is visible form PDOS and COHP diagrams. PDOS panel presents Au states from both layers. They are essentially identical which confirms the efficient screening of the field in the metal. The other two lines in PDOS panel represent the states of two Ga-N atom pairs. The first pair Ga-N1 represent the DOS of the pair of Ga-N atoms in the second layer from the top. These atoms are bond only to other Ga and N atoms, so their states are not affected by bonding to Au. The second pair Ga-N2 is located on the second layer from the bottom. These states are bound to other Ga and N atoms as well. Thus their states are not affected by bonding to hydrogen pseudoatoms. The observed energy difference of about 1.2 eV is related to the potential profile within GaN slab. The shift is confirmed by COHP plots. In addition, COHP plots show peak related to the positive (i.e. bonding) overlap of Au-and Ga interface atoms, located at the Fermi energy. This confirm good wetting of Au and GaN which is related to creation of the bond between these layers.

The magnitude of the shift, i.e. electric potential profile difference $\Delta V$ may be estimated using energy DOS maxima in Au and GaN separate and in heterostructure. In separate systems these were $E_{DOS}^{DFT}(Au) = -6.26\ eV$ and $E_{DOS}^{DFT}(GaN) = -8.15\ eV$. The heterostructure PDOS maximum for GaN is diffuse due to the potential slope in GaN region. Nevertheless, the steep increase of DOS in GaN starts for the energy $\Delta E \cong 0.4\ eV$ higher than for Au. From this data we obtain the potential difference to

$$q\Delta V = E_{DOS}^{DFT}(Au) - E_{DOS}^{DFT}(GaN) - \Delta E \cong -2.29\ eV \approx -2.3\ eV \quad (2)$$

Thus, the potential difference is about $\Delta V = 2.3\ V$. This difference is partially attributed to Fermi level in the system which is located about 0.6 eV above VBM. In Mg-doped p-type GaN Fermi level is located about 0.18 eV above VBM. This adds additional contribution to the potential difference which could be estimated to be $\Delta V \approx 2.6\ V$. The potential difference $\Delta V_0 = 2.0\ V$ will be used as benchmark for further investigations.

It has to be added, that this potential difference may be partially affected by the doping in p-type GaN. Similar simulations for other doping levels leads to the Fermi level located in different energies: (i) in upper part of bandgap for $N_A = 10^{19}\ cm^{-3}$, (ii) at VBM for $N_A = 10^{21}\ cm^{-3}$. Naturally, the Fermi level position is affected by the volume of bulk Au and p-GaN. Further increase of Au volume leads to increase of the Fermi energy while higher p-type doping leads to the downward shift. The obtained estimate and the detailed ab initio picture confirm basic assumptions of drift-diffusion heterojunction model which will be used below for design of new type contact.

### c. Heterojunction potential profiles

In general, the three possible metal-semiconductor heterojunction types were identified [12]. They were proposed by Schottky and Mott [49,50], Bardeen and Heine [51,52] and Tersoff [53]. The models differ by the assumptions of the influence of interface states which may pin the Fermi level. Since such states are not confirmed, we adopted the simplest assumption reducing to Schottky-Mott model, given by Eq. 1. In this paper, we propose the use of this model to create a controlled defect structure of the contact via multiple implantation of the nitride surface sublayers. These sublayers should be implanted across the entire band-bended region. First, the high density of Mg acceptors shall be introduced. According to the high concentration Mg doping during growth, the Mg acceptor density $N_A = 5 \times 10^{19}\ cm^{-3}$ is used. Higher density may lead to creation of Mg-Mg pairs which is detrimental to p-type doping.

The potential profiles are obtained from uniaxial Poisson equation for electric potential V(z):

$$\frac{d^2V}{dz^2} = \frac{e[N_A^- - N_D^+ + n - p]}{\varepsilon_{GaN}\varepsilon_o} \tag{3}$$

where e is elementary charge, $\varepsilon_{GaN}$ – GaN dielectric constant, $\varepsilon_o$ – vacuum permittivity and $N_A^-, N_D^+, n, p$ – densities of charged acceptors, donors, electron and holes, respectively. Neutral defects do not contribute to this equation. The simplified Schottky picture considers fully ionized acceptors only, i.e. $N_A^- = const$ and $N_D^+ = n = p = 0$. Then the solution is simple parabolic dependence ($0 \leq z \leq z_o$)

$$V(z) = V_o + \frac{eN_A^-[z-z_o]^2}{2\varepsilon_{GaN}\varepsilon_o} \qquad (4)$$

where $L_{Sch} = z_o$ – is the Schottky width of charged layer and $V_o$ – far distance potential value. From the potential deviation (jump) amplitude $\Delta V_o$ the Schottky width may be obtained as $L_{Sch} = \sqrt{\frac{2\varepsilon_{GaN}\varepsilon_o \Delta V_o}{eN_A}}$ [23]. The above specified density of acceptors density $N_A^- = 10^{19} \, cm^{-3}$ and the potential difference $\Delta V_o \cong 2 \, eV$ give the width $L_{Sch} = 4.70 \times 10^{-7} m = 4.70 \times 10^3 \text{Å}$. This width cannot be overcome by direct tunneling.

More precise description is obtained by taking into account presence of defects and mobile carriers, i.e. solving full Eq. 3 reformulated as:

$$\frac{d^2v(u)}{du^2} = \frac{[N_A^-(u) - N_A^+(u) + n(u) - p(u)]}{(p_b + n_b)} \qquad (5)$$

using dimensionless quantities: thermal energy scaled potential $v \equiv \frac{e(V-V_o)}{kT}$ ($k$ – Boltzmann constant), length $u \equiv \frac{z}{L_D}$, where Debye-Hückel screening length is $L_D \equiv \left[\frac{kT\varepsilon_o\varepsilon_{GaN}}{e^2(p_b+n_b)}\right]^{1/2}$ [23]. For the benchmark temperature, $T = 300 \, K$, the screening length extends from $L_D = 3.82 \times 10^{-6} \, m$ to $L_D = 1.21 \times 10^{-7} \, m$, respectively. This is wider that Schottky length ($L_D > L_{Sch}$) which is understandable as the latter assumes complete ionization of acceptors. Since the charge is screened by electrons and holes in the bulk of the density $n_b$ and $p_b$, respectively, the far distance limit is $\lim_{u \to \infty} v = 0$. The far distance values are used as the reference energies, therefore the energy of the conduction/valence band are shifted by the potential as $E_{C,V}(z) = E_{C,V}(\infty) - eV(z)$. Inside the layer the carrier density is position dependent, that can be expressed using scaled potential as:

$$n(u) = \frac{2N_C}{\sqrt{\pi}} F_{1/2}(\eta_F - \eta_C(u)) = \frac{2N_C}{\sqrt{\pi}} F_{1/2}(\eta_F - \eta_C + v) \qquad (6a)$$

$$p(u) = \frac{2N_V}{\sqrt{\pi}} F_{1/2}(\eta_V(u) - \eta_F) = \frac{2N_V}{\sqrt{\pi}} F_{1/2}(\eta_V - v - \eta_F) \qquad (6b)$$

where $F_j(x) \equiv \int_0^\infty \frac{y^j dy}{1+exp(y-x)}$ in Fermi integral and $N_{V,C} = 2M_{C,V}\left(\frac{2\pi m_{e,h}^* kT}{h^2}\right)^{3/2}$ are effective densities of states in conduction (C) and valence (V) bands, respectively. The factor is introduced to take into account the value of Fermi integral at zero $F_{1/2}(0) = \int_0^\infty \frac{y^{1/2}dy}{1+exp(y)} = \frac{\sqrt{\pi}}{2}$, thus $n(0) = N_C$ and $p(0) = N_V$. The $M_{C,V}$ coefficients denote the number of branches in conduction and valence bands, effectively participating in the screening, i.e. $M_C = 1$ and $M_V = 3$. The other symbols are standard: $m^*_{e,h}$ is electron/hole effective mass, h – Planck constant. The dimensionless energies (at far distance) are denoted as: $\eta_F = \frac{E_F}{kT}$ – Fermi energy, $\eta_C = \frac{E_C}{kT}$ – conduction band minimum (CBM), $\eta_V = \frac{E_V}{kT}$ - valence band maximum (VBM).

Typically the energy of the defect states (donors and acceptors) is shifted by electric potential in the same way as the change of the band energy, i.e. $E_{D,A}(z) = E_{D,A}(\infty) - eV(z)$. The charged defects contribute to the screening, thus their densities are:

$$N_D^+(u) = \frac{N_D}{1+exp[\eta_F-\eta_D(u)]} = \frac{N_D}{1+exp[\eta_F-\eta_D+v(u)]} \tag{7a}$$

$$N_A^-(u) = \frac{N_A}{1+exp[\eta_A(u)-\eta_F]} = \frac{N_A}{1+exp[\eta_A-v(u)-\eta_F]} \tag{7b}$$

where $N_{A,D}$ -are the total densities of donors/acceptors, respectively.

The fundamental equation for the potential in the charged layer in the dimensionless form is [54,55]:

$$\frac{d^2v(u)}{du^2} = \frac{1}{(n_b+p_b)} \left\{ \frac{N_A}{[1+2exp(\eta_A-\eta_F-v)]} - \frac{N_D}{[1+2exp(\eta_F-\eta_D+v)]} + n(\eta_C - \eta_F - v) - p(\eta_F - \eta_V + v) \right\} \tag{8}$$

which can be reformulated into the form suitable for wide bandgap semiconductors:

$$\frac{d^2v(u)}{du^2} = \frac{1}{(n_b+p_b)} \left\{ \frac{N_A}{[1+2exp(\eta_A-\eta_F-v)]} - \frac{N_D}{[1+2exp(\eta_F-\eta_A+v)]} + \frac{2}{\sqrt{\pi}} [N_C F_{1/2}(\eta_F - \eta_C + v) - N_V F_{1/2}(\eta_V - \eta_F - v)] \right\} \tag{9}$$

Eq. 9 is integrable directly by multiplication by $\frac{dv(u)}{du}$ and straightforward integration to determine the magnitude of electric field $|\vec{E}_z|$ [54,55]:

$$\frac{kT|\vec{E}_z|}{e} = \frac{dv(u)}{du} = 2 \left\{ \begin{array}{l} \frac{N_A}{n+p} ln\left[\frac{2+exp(\eta_F-\eta_A)}{2+exp(\eta_F-\eta_A-v)}\right] + \frac{N_D}{n+p} ln\left[\frac{2+exp(\eta_D-\eta_F+v)}{2+exp(\eta_D-\eta_F)}\right] \\ + \frac{4N_C}{3(n+p)\sqrt{\pi}} [F_{3/2}(\eta_F - \eta_C + v) - F_{3/2}(\eta_F - \eta_C)] \\ + \frac{4N_V}{3(n+p)\sqrt{\pi}} [F_{3/2}(\eta_V - \eta_F - v) - F_{3/2}(\eta_V - \eta_F)] \end{array} \right\}^{1/2} \tag{10}$$

This expression is not amenable for effective analysis. Therefore the presence of minority carriers could be neglected because, at $T = 300\ K$ the carrier density $n = \frac{2N_C(T)}{\sqrt{\pi}} F_{1/2}\left(-\frac{E_C - E_F}{kT}\right) = 1.42 \times 10^{-37}\ cm^{-3}$ i.e. effectively $n \approx 0$. Thus simplified expression for single mobile carrier screening is:

$$\frac{kT|\vec{E}_z|}{e} = \frac{dv(u)}{du} = 2 \left\{ \begin{array}{l} \frac{N_A}{p} \ln\left[\frac{2 + exp(\eta_F - \eta_A)}{2 + exp(\eta_F - \eta_A - v)}\right] + \frac{N_D}{p} \ln\left[\frac{2 + exp(\eta_D - \eta_F + v)}{2 + exp(\eta_D - \eta_F)}\right] \\ + \frac{4N_V}{3p\sqrt{\pi}} \left[F_{3/2}(\eta_V - \eta_F - v) - F_{3/2}(\eta_V - \eta_F)\right] \end{array} \right\}^{1/2} \quad (11)$$

In order to get electric potential, this expression should be integrated, which is not possible analytically. It has to be added that incorporation of several types of defects is possible. Since the charge at these defects is additive, this requires mere summation over all defects in the system

In order to obtain the insight into the change of effective potential the numerical integration was applied. In order to do so, the numerical values of the parameters at $T = 300\ K$ ($kT = 25.85\ meV$) were used as follows: $N_C(300\ K) = 2.21 \times 10^{18}\ cm^{-3}$, $N_C(300\ K) = 1.33 \times 10^{19}\ cm^{-3}$. For p-type doping Mg acceptor was used in the concentrations in the experimentally available range $N_A \in [10^{15}\ cm^{-3}, 1.15 \times 10^{17}\ cm^{-3}, 10^{19}\ cm^{-3}]$. The potential profiles were obtained for three landmark concentrations of acceptors: $N_A = 10^{15}\ cm^{-3}$, $N_A = 1.15 \times 10^{17}\ cm^{-3}$ and $N_A = 10^{19}\ cm^{-3}$. For the calculations Schottky approximation was used, assuming the total occupation of acceptors in which the potential profile is readily obtained

$$V(z) = \frac{N_A}{2\varepsilon_{GaN}\varepsilon_o} (z - L_{Sch})^2 \quad (12)$$

The second diagram of Fig. 7 presents full dependence obtained by numerical integration of the electric field obtained in Eq. 11.

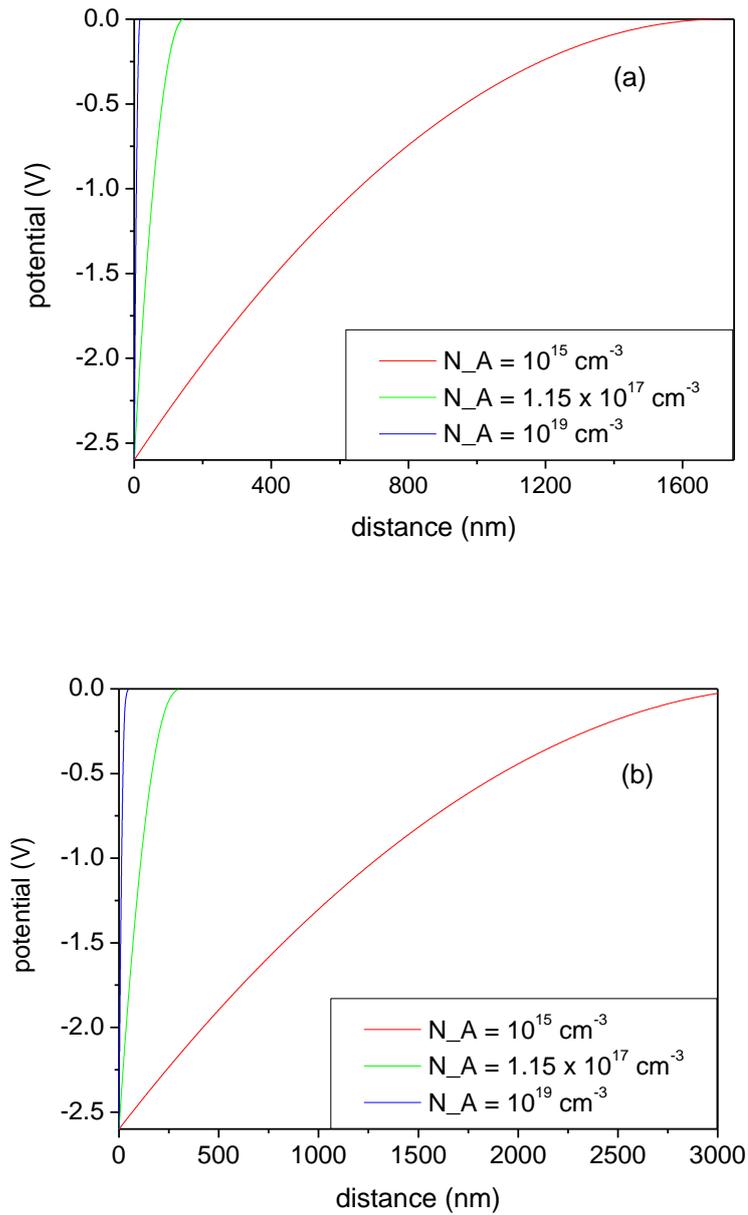

*Fig. 7. Potential profiles obtained for three selected densities of Mg acceptors: (a) Schottky approximation, (b) full solution.*

These data indicate that the Schottky approximation provides incorrect dependence, especially for higher concentration of acceptors, such as $N_A = 10^{19}\ cm^{-3}$. This could be understood from Fig. 2 as the Fermi level position is higher for lower concentration of acceptors. For concentration above $N_A = 10^{17}\ cm^{-3}$ the level falls below acceptor level thus its occupation is considerably below unity and the Schottky approximation gives large error.

For investigation of the compensation, Si donor concentration was used with the three compensation levels $N_D/N_A = 0.0$, $N_D/N_A = 0.1$ and $N_D/N_A = 0.5$. GaN bandgap was assumed to $E_g(GaN) = 3.47\ eV$ [41,42]. The bonding energy of shallow silicon donor is assumed to $20\ meV$, i.e. $E_C - E_D = 0.02\ eV$ [41]. As shown in Fig. 2 relatively small compensation affects the Fermi level for $N_A = 10^{18}\ cm^{-3}$. Accordingly, the Fermi level is close to acceptor level. Therefore the density of holes is stationary, $p \cong 2\ x\ 10^{16}\ cm^{-3}$ and $p \cong 10^{17}\ cm^{-3}$ for compensation 10% and 30% respectively. This is far below the number of charged acceptors. In order to estimate the technically important role of compensation, the potential profile obtained for the upper concentration of acceptors $N_A = 10^{19}\ cm^{-3}$ was plotted in Fig. 8.

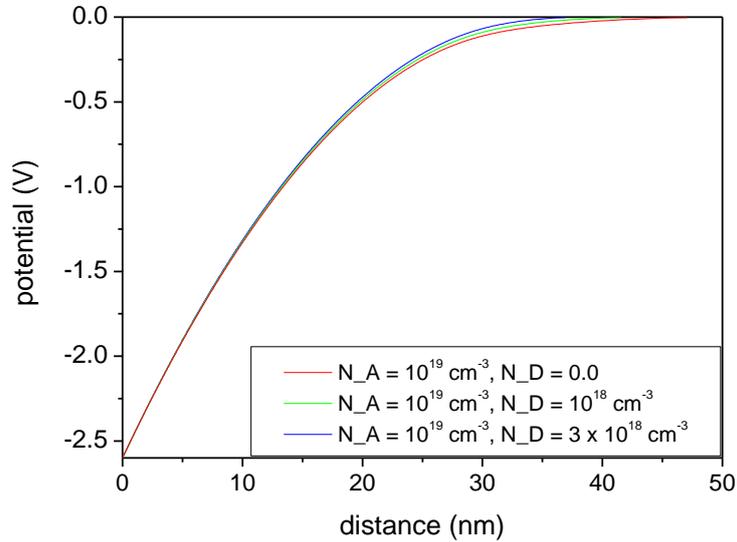

*Fig. 8. Potential profiles obtained for upper density of Mg acceptors $N_A = 10^{19}\ cm^{-3}$, obtained for three selected values of compensation: 0, 10% and 30%.*

As it is shown the influence of the compensation is essentially negligible, it extends the width of potential well by a few percent. Thus the limited presence of the donors could be neglected in the design of p-type contact.

The results of these *ab initio* and drift-diffusion calculations may be verified by direct measurements. The potential well may be measured using phase-space ab-initio direct and reverse ballistic electron emission spectroscopy [56]. In addition, combined stead-state photocapacitance (SSPC) and deep level transient spectroscopy methods open the possibility of direct determination of the deep-level trap energies and charge state [57]. For these measurement Ni electrode can be used, opening possibility of the experimental confirmation of the obtained results.

### d. Tunneling path - doping in the potential well

The extension of the potential well at the upper Mg acceptor density $N_A = 10^{19}\ cm^{-3}$ could be estimated to be $L_{well} \approx 2.5 \times 10^{-8} m = 250$ Å. This distance is about two orders of magnitude longer that is the noticeable extent of wavefunctions of the deep acceptor level. Note that shallow acceptor level are not known for gallium nitride and they cannot be used. Thus, the direct tunneling is not technically viable way of contact implementation. This is confirmed by the performance of Au as deposited contact. The annealing of Ni-Au contact in oxygen atmosphere was investigated from the metal side [10,11]. The semiconductor side was not investigated precisely. Nevertheless, it was suspected that the outdiffusion of Ga may be beneficial for the contact performance. In addition Ni defects may play role of deep acceptors with the acceptor energy level of about 1.5 – 1.6 eV above valence band maximum (VBM) [25,26]. Thus these defects could play a role in electron tunneling also.

The outdiffusion of the Ga atoms is supported by the motion of Ga vacancies. For the temperature of annealing in the range $T_{anneal} \cong 400 \div 500\ °C$ the vacancy mechanism is the only possible. Investigation of Ga vacancies were conducted for GaN by positron annihilation [58,59]. From these investigation it follows that the concentration of Ga vacancies is high in n-type GaN while in p-type materials is much reduced [60,61]. On the other hand *ab initio* investigations of Ga diffusion gave the energy barrier relatively high [62,63]. From these data it follows that creation of Ga vacancy layer is difficult, nevertheless some vacancies may be created. Therefore, it may be assumed that it is possible to create Ga vacancies to the surface. From *ab initio* investigations, published by Van de Walle and Neugebauer, it follows that Ga vacancy is deep triple acceptor [60,61]. Gallium vacancies are multiple-ionized state deep acceptors, of three ionization energies ($E_A(V_{Ga}^{1-}) \cong 2.35\ eV$, $E_A(V_{Ga}^{2-}) = 1.02 \div 0.62\ eV$, and $E_A(V_{Ga}^{3-}) \cong 0.48\ eV$) [24,25]. Thus the most upper level the initial step in the path. The two other are located lower, thus they could create a ladder with the barrier energy of order of $\Delta E_{bar} \approx 1.0\ eV$, i.e. considerably lower that the entire barrier. These Ga vacancies may serve as path for electron tunneling. Even if their location and concentration is not precisely controlled, the conduction effect may be considerable.

From the above obtained results it follows that creation of defect layer in controlled manner is indispensable for good conductivity contact to p-type GaN. The solution is to use acceptors of the energy levels above VBM in the following range $E_A \in [0.2\ eV, 2eV]$. Fortunately, extensive data on deep acceptors are available. The deepest state may be Ga vacancy with the energy $E_A(V_{Ga}) \cong 2.3 \div 2.0\ eV$. The next in ladder may be already used Ni acceptor with the energy $E_A(Ni) \cong 1.6 \div 1.5\ eV$ [36,37]. Next in the ladder may be Mn located in Ga site with the energy level at $E_A(Mn_{Ga}) \cong 1.42\ eV$ [64]. Carbon on nitrogen site has the ionization energy indicating that could be used as next in the ladder $E_A(C_{Ga}) \cong 0.9 \div 1.1\ eV$ [65]. This acceptor may be followed by mercury with $E_A(Hg_{Ga}) \cong 0.77\ eV$ used and cadmium which has estimated ionization energy at about $E_A(Cd_{Ga}) \cong 0.55\ eV$ [66,67]. Finally beryllium could be used with much smaller ionization energy at $E_A(Be_{Ga}) \cong 0.35\ eV$ [67,68] towards substitutional Mg at $E_A(Mg_{Ga}) \cong 0.18\ eV$ [3,67]. In addition, iron can be used at the first stage as it is frequently incorporated during growth as unintentional doping with the energy level at $E_A(Mg_{Ga}) \cong 2.6\ eV$ above VBM [65,66]. The design of the new low resistance contact to p-type GaN may be constructed using these data. Such design is presented in Fig. 9.

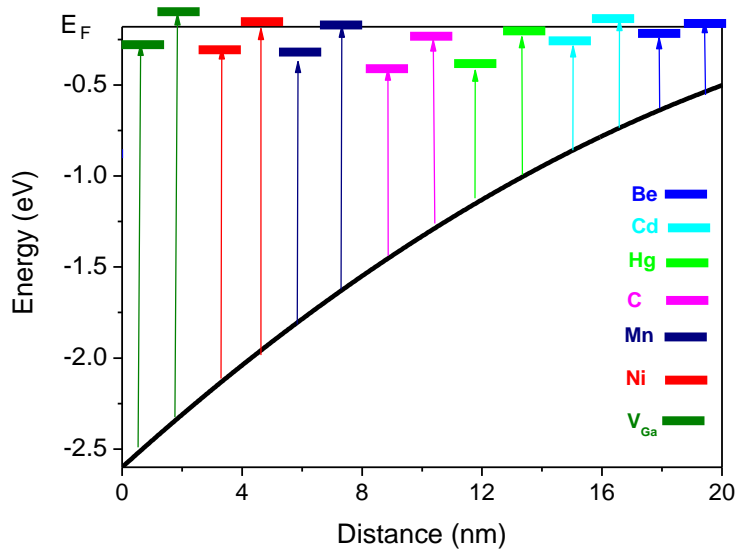

*Fig. 9. Design of multilayer contact to p-type GaN with the maximal density of Mg acceptors $N_A = 10^{19}\ cm^{-3}$ based on calculated potential profile. The horizontal color bars represent the dopant acceptor levels. The arrows represent the acceptor ionization energy.*

From this design it follows that the closest part of the multilayer contact is similar to the naturally formed contact. The additional part has to added by implantation of the additional atoms. The depth of the layer may be determined from this diagram

i/ $V_{Ga}$ $\quad\quad z \in (0,\ 2.5\ nm)$

ii/ Ni $\quad\quad z \in (2.0\ nm,\ 6.0\ nm)$

iii/ Mn $\quad\quad z \in (5.0\ nm,\ 9.0\ nm)$

iv/ C $\quad\quad z \in (8.0\ nm,\ 12.0\ nm)$

v/ Hg $\quad\quad z \in (11.0\ nm,\ 15.0\ nm)$

vi/ Cd $\quad\quad z \in (14.0\ nm,\ 17.0\ nm)$

vii/ Be $\quad\quad z \in (16.0\ nm,\ 25.0\ nm)$

The implanted layers have to be annealed before deposition of Ni contact. The Ni and $V_{Ga}$ layer may be added by standard formation of Ni/Au contact, i.e. annealing in oxygen atmosphere. If possible, the alternative procedure of creation of these layer may be used. In addition, iron may be added, depending on the type of the metal used. It has to be mentioned that this is rough estimate, in fact much wider region could be used, or even the entire well thickness could be implanted by all dopants. This has to be used with some caution as the ionized acceptors may impede the electron tunneling. On the other hand it is much easier technically.

It has to be added that these energies are burdened by some errors so the precise implementation of these contact will require series of experiments. Also, the discovery of additional acceptors may be useful in fine tuning of the contact. On the other hand, stability of such structure may be enhanced due to absence of standard contact formation in case when the alternative method is devised. It would be also useful substitution of Ga vacancy by other deep substitutional acceptor as such structure would be much more stable.

Finally, the design in Fig. 9 is presented for explanation of the idea. Naturally, these layers can overlap. In fact, the accepter with smaller activation energies may be implanted in lower depth, e.g. beryllium can be implanted over the width of the contact barrier. Those which are closer to the surface will be occupied so they would play the role of deep acceptor, i.e. the $Be^-$ charge will add to $Mg^-$ charge in limiting the width of the well. The opposite, i.e. implanting Ni far away from the heterointerface will not be useful. These defect states will be far above the Fermi level, so they will not be occupied. Therefore deep implantation of the high energy defect should be avoided.

## V. Summary

The results obtained in this paper can be assessed best by representing (i) the state of the art before publication, (ii) resume of the results in the paper (iii) the state of the art after publication.

Accordingly, the state of the are before publication was:

(i) Any contact to p-type GaN has high resistance, which hampers the development of the devices seriously,

(ii) The best contact is deposited Ni/Au double layer and short time formation in oxygen atmosphere in temperature above $400\ °C$ ($T > 400\ °C$),

(iii) The formation leads to diffusion of Ni across Au layer to create NiO,

(iv) The influence of the contact formation on the semiconductor part is not available .

The resume of the results obtained within this publication:

(i) Metal – p-type GaN contact leads to transfer of electrons from metal to semiconductor part, creation of dipole layer, and equality of the Fermi level in the system,

(ii) *Ab initio* investigation of Au-GaN heterostructure band diagram confirm electron transfer from Au to p-type GaN,

(iii) The potential profile obtained from *ab initio* investigation of Au-GaN heterostructure indicate that the depth of the potential well is higher than 2V,

(iv) The thickness of the potential well decreases rapidly for higher concentration of Mg acceptors, for concentration density $N_A = 10^{19}\ cm^{-3}$ could be estimated to be $L_{well} \approx 2.5 \times 10^{-8} m = 250\ Å$,

(v) The thickness of the potential well does not depend on the donor compensation up to the relatively high (30%) compensation level,

(vi) The thermal annealing formation of Ni/Au contact leads to outdiffusion of Ga and influx of Ni thus so created defects $V_{Ga}$ and Ni provide quantum states necessary for the electron tunneling, the main manner of holes influx into p-type GaN

(vii)   The multilayer contact structure, created by controlled implantation of the well-defined set of deep acceptors could create effective tunneling path, necessary for the low resistance contact.

The state of the art after the publication:

(i)   The semiconductor part of the structure of the metal – p-type GaN Ni/Au contact was elucidated,

(ii)   The electron transfer from metal to p-type GaN is essential part of Au-GaN contact as confirmed by *ab initio* calculations

(iii)   The dependence of the potential well width on the acceptor doping level was established,

(iv)   The compensation(up to 30%) plays negligible role in the contact properties,

(v)   The design of new, possibly low-resistance multilayer contact, to p-type GaN was proposed.

These results pave the way to faster development of wide class of nitride optoelectronic devices, both LEDs and LDs.

**Acknowledgements**

This research was carried out with the support of the Interdisciplinary Centre for Mathematical and Computational Modelling at the University of Warsaw (ICM UW) under grant nos. GB84-23 and GB99-2112.

**References**

[1]   I. Akasaki, H. Amano, S. Nakamura, Nobel Prize 2014.
      https://www.nobelprize.org/prizes/physics/2014/summary/

[2]   S. Nakamura, G. Fasol, S. J. Pearton, The Blue Laser Diode: The Complete Story, Springer, Berlin 2000.

[3]   E. Litwin-Staszewska, R. Piotrzkowski, L. Dmowski, P. Prystawko, R. Czernecki, L. Konczewicz, J Appl. Phys. 99, 03703 (2006). https://doi.org/10.1063/1.2168232

[4]   J. Simon, V. Protasenko, C. Lian, H. Xing, D. Jena, Science 327, 60 (2010).
      https://doi.org/10.126/science.1183226

[5]   A. Ahmad, P. Strak, P. Kempisty, K. Sakowski, J. Piechota, Y. Kangawa, I. Grzegory, M. Leszczynski, Z. R. Zytkiewicz, G. Muziol, E. Monroy, A. Kaminska, S. Krukowski, J. Appl. Phys. 132, 064301 (2022). https://doi.org/10.1063/5.0098909

[6]     Y. Sato, K. Yamada, K. Sakowski, M. Iwaya, T. Takeuchi, S. Kamiyama, Y. Kangawa, P. Kempisty, S. Krukowski, J. Piechota, Appl. Phys. Expr. 14, 0965903 (2021). https://doi.org/10.35848/1882-0786/ac1d64

[7]     M. Aktas, S. Grzanka, L.Marona, J. Goss, G. Staszczak, P. Perlin, Materials 17, 4502, (2024). https://doi.org/10.3390/ma17184502

[8]     CRC Handbook on Chemistry and Physics, 2008, p. 12-24.

[9]     L. -C. Chen, F. -R. Chen, J.-J. Kai, L. Chang, J.-K. Ho, C. -S. Jong, C. C. Chiu, C.-N. Huang, C.-Y. Chen, K.-K Shih, Microstructural investigation of oxidized Ni/Au ohmic contact to *p*-type GaN. J. Appl. Phys. 86, 3826 (1999). https://doi.org/10.1063/1.371294

[10]    J. Smalc-Koziorowska, S. Grzanka, E. Litwin-Staszewska, R. Piotrzkowski, G. Nowak, M. Leszczynski, P.Perlin, E. Talik, J. Kozubowski, S. Krukowski, Ni-Au contacts to p-type GaB – structure and properties. Solid State Electron. 54, 701 (2010). http://dx.doi.org/10.1016/j.sse.2010.01.026

[11]    G. Greo, P. Prystawko, M. Leszczynski, R. Lo Nigo, V. Rainieri, F. Roccaforte, Electro-structural evolution and Schottky barrier height in annealed Au/Ni contacts onto p-GaN. J. Appl. Phys. 110, 123703 (2011). https://doi.org/10.1063/1.3669407

[12]    J. -S. Jang, I. -S. Chang, H. -K. Kim, T. -Y. Seong, S. lee, S. -J. Park, Low-resistance Pt/Ni/Au ohmic contact to p-type GaN. Appl. Phys. Lett. 74, 70 (1999) . https://doi.org/10.1063/1.123954

[13]    L. Zhou, W. Landford, A. T. Ping, I. Adesida, J. W. Yang, A. Khan, Low-resistance Ti/Pt/Au ohmic contact to p-type GaN. Appl. Phys. Lett. 76, 3453 (2000). https://doi.org/10.1063/1.126674

[14]    J. K. Kim, J. L. Lee, J. W. Lee, H. E. Shin, Y. J. Park, T. Kim, Low resistance Pd/Au ohmic contacts to p-type GaN using surface treatment. Appl. Phys. Lett. 73, 2953 (1998). https://doi.org/10.1063/1.122641

[15]    P. Strak, P. Kempisty, K. Sakowski, S. Krukowski J. Vac. Sci. Technol. A 35, 021406 (2017). http://dx.doi.org/10.1116/1.4975332

[16]    Y. Koide, T. Maeda, T. Kawakami, S. Fujita, T. Uemura, N. Shibata, M. Murakami, Effects of annealing in an oxygen ambient on electrical properties of ohmic contacts to p-type GaN. J. Electron. Mater. 28, 341-346 (1999). https://doi.org/10.1007/s11664-999-0037-7

[17] L.-C. Chen, J.-K. Ho, C.-S. Jong, C.-C. Chiu, K-K Shih, F.-R. Chen, L. Chang, Oxidized Ni/Pt and Ni/Au ohmic contacts to p-type GaN. Appl. Phys. Lett. 76, 3703-3706 (2000). https://doi.org/10.1063/1.126624

[18] T. Maeda, Y. Koide, M. Murakami. Effects of NiO on electrical properties of NiAu-based ohmic contacts for p-type GaN. Appl. Phys. Lett. 75, 4145-4147 (1999). https://doi.org/10.1063/1.125564

[19] J. Weg, J. Hu, C. Guan, S. Fang, Z. Wang, G. Wang, K. Xu, T. Lv, X. Wang, J. Shi, Z. Li, J. Zhang, N. Chi, C. Shen, Low-resistance Ohmic contact for GaN-based laser diodes, J. Semicond. 45, 122502 (2024). https://doi.org/10.1088/1674-4926/24060018

[20] H. Morkoc, Metal Contact to GaN and Processing, in Handbook of Nitride Semiconductors and Devices, vol. 2. Electronic and Optical Processes in Nitrides, Wiley-VCH Weinheim 2008. p. 1-121

[21] A. Franciosi, C. G. Van de Walle, Surf. Sci. Rep. 25 (1996) 1. https://doi.org/10.1016/0167-5729(95)00008-9

[22] B. S. Eller, J. Yang, R.J. Nemanich, Electronic surface and interface states on GaN and AlGaN, J. Vac. Sic. Technol. A 31, 050807 (2013). https://doi.org/10.1116/1.4807904

[23] W. Mönch, Semiconductor Surfaces and Interfaces. Springer Berlin 1993

[24] M. Meneghini, C. De Santi , I. Abid, M. Buffolo, M. Cioni, R. A. Khadar, L. Nela, N. Zagni, A. Chini, F. Medjdoub, G. Meneghesso, G. Verzellesi, E. Zanoni, E. Matioli, GAN-based power devices: Physics, reliability and perspectives. J. Appl. Phys. 130, 181101 (2021). https://doi.org/10.1063/5.0061354

[25] C. G. Van de Walle, J. Neugebauer, First-principles calculations for defects and impurities: Applications to III-nitrides. J. Appl. Phys. 95, 3851 – 3879 (2004). https://doi.org/10.1063/1.1682673

[25] M. Rychetsky, I. L. Koslow, T. Wernicke, J. Rass, V. Hoffmann, M. Weyers, M. Kneissl, Impact of acceptor concentration on resistivity of Ni/Au p-contacts on semipolar (20-21) GaN:Mg. Phys. Stta. Sol. (b) 253, 169 (2016). https://doi.org/10.1002/pssb.201552407

[26] T. Odani, K. Iso, Y. Oshima, H. Ikeda, H. Ikeda, T. Mochizuki S. Iuzmisawa, Realization of High-Resistive Ni-doped GaN Crystal by Hydride Vapor Phase Epitaxy, Phys. Stat. Sol. (b) 261, 2300584 (2024) . https://doi.org/10.1002/pssb.202300584

[27] A. Garcia et al. SIESTA: Recent developments and applications. J. Chem. Phys. 152, 204108 (2020). https://doi.org/10.1063/5.0005077

[28] J. Junquera, O. Paz, D. Sanchez-Portal, E. Artacho, Numerical atomic orbitals for linear-scaling calculations. Phys. Rev. B **64**, 235111 (2001). https://doi.org/10.1103/PhysRevB.64.235111

[29] E. Anglada, J. M. Soler, J. Junquera, E. Artacho, Systematic generation of finite-range atomic basis sets for linear-scaling calculations. Phys. Rev. B 66, 205101 (2002). https://doi.org/10.1103/PhysRevB.66.205101

[30] R. Coquet, G. J. Hitchings, S. H. Taylor, D. J. Willock, Calculations on the adsorption of Au to MgO surfaces using SIESTA. J. Mater. Chem. 16, 1978 – 1988 (2006). https://doi.org/10.1039/b601213b

[31] I. G. Gurtubay, J. M. Pitarke, I. CAmpillo, A. Rubio, Dynamic structure factor of gold. Comp. Mater. Sci. 22 123-128 (2001). https://doi.org/10.1016/S0927-0256(01)00178-1

[32] N. Troullier, J. L. Martins, Efficient pseudopotentials for plane-wave calculations. Phys. Rev. B, **43,** 1993-2006. (1991). https://doi.org/10.1103/PhysRevB.43.1993

[33] N. Troullier, J. L. Martins, Efficient pseudopotentials for plane-wave calculations. II. Operators for fast iterative diagonalization, Phys. Rev. B, **43,** 8861-8869 (1991) . https://doi.org/10.1103/PhysRevB.43.8861

[34] H.J Monkhorst, J. D. Pack, Special points for Brillouin-zone integrations. Phys. Rev. B, 13, 5188 - 5192 (1976). https://doi.org/10.1103/PhysRevB.13.5188

[35] L. S. Pedroza, A. J. R. da Silva, K. Capelle, Gradient-dependent density functionals of the Perdew-Burke-Ernzerhof type for atoms, molecules, and solids, Phys. Rev. B 79, 201106(R) (2009). https://doi.org/10.1103/PhysRevB.79.201106

[36] M. M. Odashima, K. Capelle, S. B. Trickey, Tightened Lieb-Oxford Bound for Systems of Fixed Particle Number, J. Chem. Theory Comput. 5, 798 - 807 (2009) . https://doi.org/10.1021/ct8005634

[37] M. Leszczynski, , H. Teisseyre, T. Suski, I. Grzegory, M. Bockowski, J. Jun, S. Porowski, K. Pakula, J.M. Baranowski, C.T. Foxon, T.S. Cheng, Lattice parameters of gallium nitride Appl. Phys. Lett. 69**,** 73 (1996). https://doi.org/10.1063/1.118123

[38] B. Dayal, Lattice constants and thermal expansion of gold up to 878º by X-ray method, Phys. Stat. Sol. 3, 473 – 477 (1963). https://doi.org/10.1002/pssb.19630030312

[39] L.G. Ferreira, M. Marques, and L.K. Teles, Approximation to density functional theory for the calculation of band gaps of semiconductors. Phys. Rev. B 78, 125116, (2008). https://doi.org/10.1103/PhysRevB.78.125116

[40] M. Ribeiro, L.R.C. Fonseca, and L.G. Ferreira, Accurate prediction of the Si/SiO2 interface band offset using the self-consistent *ab initio* DFT/LDA-1/2 method. Phys. Rev. B, 79, 241312 (2009). https://doi.org/10.1103/PhysRevB.79.241312

[41] B. Monemar, J.P. Bergman, I. A. Buyanova, H. Amano, I. Akasaki, T. Detchprohm, K. Hiramatsu, and N. Sawaki, The excitonic bandgap of GaN: Dependence on substrate. Solid State Electron. 41, 239 (1997). https://doi.org/10.1016/S0038-1101(96)00208-0

[42] Y.C. Yeo, T.C. Chong, M.F. Li, Electronic band structures and effective-mass parameters of wurtzite GaN and InN. J. Appl. Phys. 83, 1429 (1997). https://doi.org/10.1063/1.366847

[43] C. Stampfl, C. G. Van de Walle, Density-functional calculations for III-V nitrides using local-density approximation and generalized gradient approximation, Phys. Rev. B 59, 5521 – 5535 (1999). https://doi.org/101103.PhysRevB.59.5521

[44] J. -M. Wagner, F. Bechstedt, Properties of strained wurtzite GaN and AlN; Ab initio studies, Phys. Rev. B 66, 115202 (2002). https://doi.org/10.1103/PhysRevB.66.115202

[45] P. Strak, P. Kempisty, M. Ptasinska, S. Krukowski, Principal physical properties of GaN/AlN multiquantum well (MQW) systems determined by density functional theory (DFT) calculations. J. Appl. Phys. 113, 193706 (2013). http://dx.doi.org/10.1063/1.4805057

[46] Vinayak Keshavlal Patel, Lattice constants, thermal expansion coefficients, densities and imperfections in gold and the alpha-phase of the gold-indium system (1967) . Master Theses. 6876. Missouri S&T. USA. https://scholarsmine.mst.edu/masters_theses/6876

[47] R. Dronskowski, P. E. Bloechl, Crystal orbital Hamilton populations (COHP): energy-resolved visualization of chemical bonding in solids based on density-functional calculations. J. Phys. Chem. 97, 8617 (1993). https://doi.org/10.1021/j100135a014

[48] V. L. Deringer, A. L. Tchougreeff, R. Dronskowski, Crystal Orbital Hamilton Population (COHP) Analysis as Projected from Plane-Wave Basis Sets. J. Phys. Chem. 115, 5461 (2011). https://doi.org/10.1021/jp202489s

[49] W. Schottky, Halbleitertheorie der Sperrschicht- und Spitzengleichrichter, Phys. Z. 41, 570 (1940). (in German)

[50] N. F. Mott, Note on the contact between a metal and an insulator or semi-conductor, Math. Proc. Cambridge 34, 568 (1938).

[51] J. Bardeen, Surface States and Rectification at a Metal Semi-Conductor Contact, Phys. Rev. 71, 717 (1947). https://doi.org/10.1103/PhysRev.71.717

[52] V. Heine, Theory of Surface States, Phys. Rev. 138, A1689 (1965). https://doi.org/PhysRev.138.A1689

[53] J. Tersoff, Theory of Semiconductor Heterojunctions: The Role of Quantum Dipoles, Phys. Rev. B 30, 4874 (1984). https://doi.org/10.1103/PhysRevB.30.4874

[54] R. Seiwatz, M. Green, Space Charge Calculations for Semiconductors, J. Appl. Phys. 29, 034 (1958). https://doi.org/10.1063/1.1723358

[55] S. Krukowski, P. Kempisty, P. Strak, Foundations of ab initio simulations of electric charges and fields at semiconductor surfaces within slab models. J. Appl. Phys. 114, 143705 (2013). http://dx.doi.org/10.1063/1.4824800

[56] A. Gerbi, R. Buzio, C. Gonzales, F. Flores, P. L. de Andres, Phase-space ab-initio direct and reverse ballistic-electron emission spectroscopy: Schottky barriers determination for Au/Ge(100). Appl. Surf. Sci. 609, 155218 (2023). https://doi.org/10.1016/j.apsusc.2022.155218

[57] P. Kruszewski, P. Sai, A. Krajewska, K. Sakowski, Y. Ivonyak, R. JAkiela, J. Plesiewicz, P. Prystawko, Graphene Schottky barrier diode acting as a semi-transparent contact to n-GaN. AIP Adv. 14, 075312 (2024). https://doi.org/10.1063/5.0210798

[58] K. Saarinen, T. Laine, S. Kuisna, J. Nissila, P. Hautojaarvi, L. Dobrzynski, J.M Baranowski, K. Pakula, R. Stepniewski, M. Wojdak, M. Wysmolek, T. Suski, M. Leszczynski, I. Grzegory, S. Porowski, Observation of native Ga vacancies by positron annihilation, Phys. Rev. Lett. 79, 3030 (1997). https://doi.org/10.1103/PhysRevLett.79.3030

[59] H. Nykanen, S. Suihkonen, L. Kilanski, M. Sopanen, F. Tuomisto, Low energy electron beam induced vacancy activation. Appl. Phys. Lett. 100, 122105 (2012). https://doi.org/10.1063/1.3696047

[60] C. Van de Walle, J. Neugebauer, First-principles calculations for defects and impurities: Applications to III-nitrides. J. Appl. Phys. 95, 3851 (2004). https://doi.org/10.1063/1.1682673

[61] C.E. Dreyer, A. Alkauskas, J. Lyons, J.S. Speck, C. Van de Walle, Gallium vacancy complexes as a cause of Shockley-Read-Hall recombination in III-nitride light emitters, Appl. Phys. Lett. 108, 141101 (2016) https://doi.org/10.1063/1.4942674

[62] M. M. Ganchenkova, R. M. Nieminen, Nitrogen Vacancies as Major Point Defects in Gallium Nitride, Phys. Rev. Lett. 96, 196402 (2006). https://doi.org/10.1103/PhysRevLett.96.196402

[63] R. Hrytsak, P. Kempisty, E. Grzanka, M. Leszczynski, M. Sznajder, Modeling of the Point Defect Migration across the AlN/GaN Interfaces-Ab Initio Study. Materials 15, 478 (2022). https://doi.org/10.3390/ma15020478

[64] R. Y. Korotkov, J. Gregie, B. V. Wessel, Optical properties of the deep Mn acceptor in GaN:Mn. Appl. Phys. Lett. 80, 1731 (2002). https://doi.org/10.1063/1.1456544

[65] J. L. Lyons, A. Janotti, C. G. Van de Walle, Carbon impurities and the yellow luminescence. Appl. Phys.Lett. 97, 152108 (2010). https://doi.org/10.1063/1.3492841

[66] M. A. Reschikov, O. Andreiev, M. Vorobiov, D. O. Demchenko, B. McEwen, F. Shahedipour-Sandvik, Photoluminescence from $Cd_{Ga}$ and $Hg_{Ga}$ acceptors in GaN, J. Apppl. Phys. 135, 155706 (2024). https://doi.org/10.1063/5.0202741

[67] D. O. Demchenko, M. Vorobiov, O. Andreiev, M. A. Reschikov, Koopmans-tuned Heyd-Scuseria-Ernzerhof hybrid functional calculations of acceptors in GaN. Phys. Rev. B 110, 035203 (2024). https://doi.org/10.1103/PhysRevB.110.035203

[68] M. A. Reschikov, M. Vorobiov, O. Andreiev, D. O. Demchenko, B. McEwen, F. Shahedipour-Sandvik, Dual nature of the $Be_{Ga}$ acceptor in GaN: Evidence from photoluminescence. Phys. Rev. B. 108, 072202 (2023). https://doi.org/10.1103/PhysRevB.108.075202